\RequirePackage{lineno}
\documentclass[aps,prl,twocolumn,superscriptaddress,showpacs]{revtex4}
\usepackage{times}
 
\usepackage{setspace}
\usepackage{float}
\usepackage{epsfig}
\begin{document}

%\linenumbers

%\preprint{}

\begin{singlespace}

\title{Polarization Observables in Deuteron Photodisintegration below 360 MeV}



\author{J. Glister}
%\email[Corresponding author.\\Email address: ]{jglister@jlab.org}
%\thanks{}
\affiliation{Saint Mary's University, Halifax, Nova Scotia B3H 3C3, Canada}
\affiliation{Dalhousie University, Halifax, Nova Scotia B3H 3J5, Canada}

\author{G. Ron}
\affiliation{Tel Aviv University, Tel Aviv 69978, Israel}
\affiliation{Weizmann Institute of Science, Rehovot 76100, Israel.}

\author{B. W. Lee}
\affiliation{Seoul National University, Seoul 151-747, Korea}

\author{R. Gilman}
\affiliation{Thomas Jefferson National Accelerator Facility, Newport News, Virginia 23606, USA}
\affiliation{Rutgers, The State University of New Jersey, Piscataway, New Jersey 08855, USA}

\author{A. J. Sarty}
\affiliation{Saint Mary's University, Halifax, Nova Scotia B3H 3C3, Canada}

\author{S. Strauch}
\affiliation{University of South Carolina, Columbia, South Carolina 29208, USA}

\author{D. W. Higinbotham}
\affiliation{Thomas Jefferson National Accelerator Facility, Newport News, Virginia 23606, USA}

\author{E. Piasetzky}
\affiliation{Tel Aviv University, Tel Aviv 69978, Israel}

\author{K. Allada}
\affiliation{University of Kentucky, Lexington, Kentucky 40506, USA}

\author{W. Armstrong}
\affiliation{Temple University, Philadelphia, Pennsylvania 19122, USA}

\author{J. Arrington}
\affiliation{Argonne National Laboratory, Argonne, Illinois 60439, USA}

\author{H. Arenh\"{o}vel}
\affiliation{Institut f\"{u}r Kernphysik, Johannes Gutenberg-Universit\"{a}t, D-55099 Mainz, Germany}

\author{A. Beck}
\affiliation{NRCN, P.O. Box 9001, Beer-Sheva 84190, Israel}

\author{F. Benmokhtar}
\affiliation{University of Maryland, Baltimore, Maryland, USA}

\author{B. L. Berman}
\affiliation{George Washington University, Washington, D.C. 20052, USA}

\author{W. Boeglin}
\affiliation{Florida International University, Miami, Florida 33199, USA}
\author{E. Brash}
\affiliation{Christopher Newport University, Newport News, Virginia 23606, USA}
\author{A. Camsonne}
\affiliation{Thomas Jefferson National Accelerator Facility, Newport News, Virginia 23606, USA}
\author{J. Calarco}
\affiliation{University of New Hampshire, Durham, New Hampshire 03824, USA}
\author{J. P. Chen}
\affiliation{Thomas Jefferson National Accelerator Facility, Newport News, Virginia 23606, USA}
\author{S. Choi}
\affiliation{Seoul National University, Seoul 151-747, Korea}
\author{E. Chudakov}
\affiliation{Thomas Jefferson National Accelerator Facility, Newport News, Virginia 23606, USA}
\author{L. Coman}
\affiliation{University of Virginia, Charlottesville, Virginia 22094, USA}
\author{B. Craver}
\affiliation{University of Virginia, Charlottesville, Virginia 22094, USA}
\author{F. Cusanno}
\affiliation{INFN, Sezione Sanit\'{a} and Istituto Superiore di Sanit\'{a}, Laboratorio di Fisica, I-00161 Rome, Italy}
\author{J. Dumas}
\affiliation{Rutgers, The State University of New Jersey, Piscataway, New Jersey 08855, USA}
\author{C. Dutta}
\affiliation{University of Kentucky, Lexington, Kentucky 40506, USA}
\author{R. Feuerbach}
\affiliation{Thomas Jefferson National Accelerator Facility, Newport News, Virginia 23606, USA}
\author{A. Freyberger}
\affiliation{Thomas Jefferson National Accelerator Facility, Newport News, Virginia 23606, USA}
\author{S. Frullani}
\affiliation{INFN, Sezione Sanit\'{a} and Istituto Superiore di Sanit\'{a}, Laboratorio di Fisica, I-00161 Rome, Italy}
\author{F. Garibaldi}
\affiliation{INFN, Sezione Sanit\'{a} and Istituto Superiore di Sanit\'{a}, Laboratorio di Fisica, I-00161 Rome, Italy}
\author{J.-O. Hansen}
\affiliation{Thomas Jefferson National Accelerator Facility, Newport News, Virginia 23606, USA}
\author{T. Holmstrom}
\affiliation{Longwood University, Farmville, Virginia, 23909, USA}
\author{C. E. Hyde}
\affiliation{Old Dominion University, Norfolk, Virginia 23508, USA}
\affiliation{Universit\'{e} Blaise Pascal / CNRS-IN2P3, F-63177 Aubi\`{e}re, France}
\author{H. Ibrahim}
\affiliation{Old Dominion University, Norfolk, Virginia 23508, USA}
\author{Y. Ilieva}
\affiliation{George Washington University, Washington, D.C. 20052, USA}
\author{C. W. de Jager}
\affiliation{Thomas Jefferson National Accelerator Facility, Newport News, Virginia 23606, USA}
\author{X. Jiang}
\affiliation{Rutgers, The State University of New Jersey, Piscataway, New Jersey 08855, USA}
\author{M. K. Jones}
\affiliation{Thomas Jefferson National Accelerator Facility, Newport News, Virginia 23606, USA}
\author{A. Kelleher}
\affiliation{College of William and Mary, Williamsburg, Virginia 23187, USA}
\author{E. Khrosinkova}
\affiliation{Kent State University, Kent, Ohio 44242, USA}
\author{E. Kuchina}
\affiliation{Rutgers, The State University of New Jersey, Piscataway, New Jersey 08855, USA}
\author{G. Kumbartzki}
\affiliation{Rutgers, The State University of New Jersey, Piscataway, New Jersey 08855, USA}
\author{J. J. LeRose}
\affiliation{Thomas Jefferson National Accelerator Facility, Newport News, Virginia 23606, USA}
\author{R. Lindgren}
\affiliation{University of Virginia, Charlottesville, Virginia 22094, USA}
\author{P. Markowitz}
\affiliation{Florida International University, Miami, Florida 33199, USA}
\author{S. May-Tal Beck}
\affiliation{NRCN, P.O. Box 9001, Beer-Sheva 84190, Israel}
\author{E. McCullough}
\affiliation{Saint Mary's University, Halifax, Nova Scotia B3H 3C3, Canada}
\affiliation{University of Western Ontario, London, Ontario N6A 3K7, Canada.}
\author{D. Meekins}
\affiliation{Thomas Jefferson National Accelerator Facility, Newport News, Virginia 23606, USA}
\author{M. Meziane}
\affiliation{College of William and Mary, Williamsburg, Virginia 23187, USA}
\author{Z.-E. Meziani}
\affiliation{Temple University, Philadelphia, Pennsylvania 19122, USA}
\author{R. Michaels}
\affiliation{Thomas Jefferson National Accelerator Facility, Newport News, Virginia 23606, USA}
\author{B. Moffit}
\affiliation{College of William and Mary, Williamsburg, Virginia 23187, USA}
\author{B. E. Norum}
\affiliation{University of Virginia, Charlottesville, Virginia 22094, USA}
\author{Y. Oh}
\affiliation{Seoul National University, Seoul 151-747, Korea}
\author{M. Olson}
\affiliation{Saint Norbert College, Greenbay, Wisconsin 54115, USA}
\author{M. Paolone}
\affiliation{University of South Carolina, Columbia, South Carolina 29208, USA}
\author{K. Paschke}
\affiliation{University of Virginia, Charlottesville, Virginia 22094, USA}
\author{C. F. Perdrisat}
\affiliation{College of William and Mary, Williamsburg, Virginia 23187, USA}
\author{M. Potokar}
\affiliation{Jo\v{z}ef Stefan Institute, 1000 Ljubljana, Slovenia}
\author{R. Pomatsalyuk}
\affiliation{Thomas Jefferson National Accelerator Facility, Newport News, Virginia 23606, USA}
\affiliation{NSC Kharkov Institute of Physics and Technology, Kharkov 61108, Ukraine}
\author{I. Pomerantz}
\affiliation{Tel Aviv University, Tel Aviv 69978, Israel}
\author{A. Puckett}
\affiliation{Massachusetts Institute of Technology, Cambridge, Massachusetts 02139, USA}
\author{V. Punjabi}
\affiliation{Norfolk State University, Norfolk, Virginia 23504, USA}
\author{X. Qian}
\affiliation{Duke University, Durham, North Carolina 27708, USA}
\author{Y. Qiang}
\affiliation{Massachusetts Institute of Technology, Cambridge, Massachusetts 02139, USA}
\author{R. D. Ransome}
\affiliation{Rutgers, The State University of New Jersey, Piscataway, New Jersey 08855, USA}
\author{M. Reyhan}
\affiliation{Rutgers, The State University of New Jersey, Piscataway, New Jersey 08855, USA}
\author{J. Roche}
\affiliation{Ohio University, Athens, Ohio 45701, USA}
\author{Y. Rousseau}
\affiliation{Rutgers, The State University of New Jersey, Piscataway, New Jersey 08855, USA}
\author{A. Saha}
\affiliation{Thomas Jefferson National Accelerator Facility, Newport News, Virginia 23606, USA}
\author{B. Sawatzky}
\affiliation{University of Virginia, Charlottesville, Virginia 22094, USA}
\affiliation{Temple University, Philadelphia, Pennsylvania 19122, USA}
\author{E. Schulte}
\affiliation{Rutgers, The State University of New Jersey, Piscataway, New Jersey 08855, USA}
\author{M. Schwamb}
\affiliation{Institut f\"{u}r Kernphysik, Johannes Gutenberg-Universit\"{a}t, D-55099 Mainz, Germany}
\author{M. Shabestari}
\affiliation{University of Virginia, Charlottesville, Virginia 22094, USA}
\author{A. Shahinyan}
\affiliation{Yerevan Physics Institute, Yerevan 375036, Armenia}
\author{R. Shneor}
\affiliation{Tel Aviv University, Tel Aviv 69978, Israel}
\author{S.~\v{S}irca}
\affiliation{Dept. of Physics, University of Ljubljana, 1000 Ljubljana, Slovenia}
\author{K. Slifer}
\affiliation{University of Virginia, Charlottesville, Virginia 22094, USA}
\author{P. Solvignon}
\affiliation{Argonne National Laboratory, Argonne, Illinois 60439, USA}
\author{J. Song}
\affiliation{Seoul National University, Seoul 151-747, Korea}
\author{R. Sparks}
\affiliation{Thomas Jefferson National Accelerator Facility, Newport News, Virginia 23606, USA}
\author{R. Subedi}
\affiliation{Kent State University, Kent, Ohio 44242, USA}
\author{G. M. Urciuoli}
\affiliation{INFN, Sezione Sanit\'{a} and Istituto Superiore di Sanit\'{a}, Laboratorio di Fisica, I-00161 Rome, Italy}
\author{K. Wang}
\affiliation{University of Virginia, Charlottesville, Virginia 22094, USA}
\author{B. Wojtsekhowski}
\affiliation{Thomas Jefferson National Accelerator Facility, Newport News, Virginia 23606, USA}
\author{X. Yan}
\affiliation{Seoul National University, Seoul 151-747, Korea}
\author{H. Yao}
\affiliation{Temple University, Philadelphia, Pennsylvania 19122, USA}
\author{X. Zhan}
\affiliation{Massachusetts Institute of Technology, Cambridge, Massachusetts 02139, USA}
\author{X. Zhu}
\affiliation{Duke University, Durham, North Carolina 27708, USA}
\collaboration{The Jefferson Lab Hall A Collaboration}
\noaffiliation

\date{\today}

\begin{abstract}
High precision measurements of induced and transferred recoil proton polarization 
in $d(\vec{\gamma},\vec{p})n$ have been performed for photon energies of 277--357 MeV 
and $\theta_{cm}$ = 20$^{\circ}$ -- 120$^{\circ}$.  The measurements were motivated 
by a longstanding discrepancy between meson-baryon model calculations and data at higher energies.  
At the low energies of this experiment, theory continues to fail to reproduce the data, 
indicating that either something is missing in the calculations and/or there is a problem 
with the accuracy of the nucleon-nucleon potential being used.
\end{abstract}

%Forces in hadronic systems and effective interactions, Polarization phenomena in reactions, Nuclear reactions involving few-nucleon systems, Photonuclear reactions 
%\keywords{}

\maketitle

\end{singlespace}

The conventional nuclear model uses baryon and meson degrees of freedom 
to describe nuclear structure and reactions.  While this approach has been 
broadly successful for low-energy phenomena, it is widely believed that it 
will break down at high energies.  Meson-baryon model (MBM) calculations 
require high precision $NN$ potentials
which are used to describe the finite spatial extent of 
hadrons~\cite{Machleidt:1987hj} and contain free parameters fit 
to experimental $NN$ scattering data.  MBM calculations have been 
quite successful below excitation energies of a few hundred MeV in describing 
cross-section and polarization observables for electromagnetic reactions 
involving small nuclear systems~\cite{Arenhoevel:2002kh,Marcucci:2005zc,Ahrenhoevel:1991pa,Carlson:1997qn}.  
  
In the few GeV energy region, cross section measurements of the deuteron 
photodisintegration reaction~\cite{Mirazita:2004rb} were found to approximately 
scale according to the constituent counting rules~\cite{Brodsky:1973kr,Matveev:1973ra,Rossi:2004qm}, 
predictions based on quark degrees of freedom.  Also, quark models such as the 
quark-gluon string (QGS)~\cite{Grishina:2002ph,Grishina:2003as} and 
hard rescattering (HR)~\cite{Sargsian:2003sz} have been moderately successful 
in describing $d(\vec{\gamma},\vec{p})n$ polarization observables 
above $\sim$ 1 GeV~\cite{Wijesooriya:2001yu,Jiang:2007ge}.

Small nuclear systems, such as the deuteron and $^{3}$He, provide a useful testing ground for MBM 
calculations as they allow for reliable theoretical calculations.  Electromagnetic probes of these 
small systems are useful since the weak coupling constant allows for perturbative methods to be used.  
Furthermore, polarization measurements in electron- and photon-deuteron reactions allow for a 
detailed study as they are sensitive to small amplitudes and effects.  Since the beginning of 
polarization measurements, over 70 publications have presented over 1200 polarization data points 
for photodisintegration and the time-reversed radiative capture reaction.  These data have been 
very useful in constraining and testing low energy MBM calculations.  In order to test the 
upper limit in energy of the meson-baryon model, experiments and calculations have been 
extended to higher and higher energies.  As the energy and momentum transfer increase, 
the distance scale probed decreases and one would expect that at some point 
the sub-nucleonic degrees of freedom would have to be considered.  

The most advanced MBM calculation for $d(\vec{{\gamma}},\vec{p})n$ 
in the few hundred MeV region comes from Schwamb and 
Arenh\"{o}vel~\cite{Schwamb:1999qd,Schwamb:2000zr,Schwamb:2001jy,Arenhoevel:2002kh}.  
They have included meson-exchange currents, final-state interactions, relativistic 
corrections and a modern baryon-baryon potential in a non-relativistic field 
theory with nucleon, meson and $\Delta$ degrees of freedom.  Free parameters 
are constrained by fits to $NN$ scattering, $\pi N$ scattering and 
pion photoproduction data~\cite{Schwamb:1999qd}.  Up to excitation energies 
of roughly 500 MeV, there is generally good agreement between their calculations 
and data for the differential cross section, the cross section asymmetry for 
linearly polarized photons ($\Sigma$) and the polarized target asymmetry ($T$).  
However, a striking disagreement emerges between 300 and 500 MeV, where the 
induced recoil proton polarization ($P_{y}$) at $\theta_{cm}$ = 90$^{\circ}$ 
is predicted to approach zero with increasing energy, yet the data grow in 
magnitude to nearly $-1$ at 500 MeV\@.  Kang {\it et al.}~\cite{kang1990} 
developed a model for $d(\vec{{\gamma}},\vec{p})n$ using a diagrammatic 
method which predicted a large magnitude of $P_{y}$ above 300 MeV\@.  They 
considered $\pi$, $\rho$, $\eta$ and $\omega$ meson exchanges and 17 
well-established nucleon and $\Delta$ resonances with a mass less than 2 GeV 
and $J \leq$ 5/2, with all resonance parameters taken from the Particle Data 
Group~\cite{Amsler:2008zzb}.  However, the calculation did not include channel 
coupling or consider final-state interactions completely (by solving the Schr\"{o}edinger 
equation with a $NN$ potential) and failed to describe the large induced polarization 
seen at $\sim$ 500 MeV\@~\cite{Wijesooriya:2001yu}.  The pre-existing data between 
300 and 500 MeV consist mainly of induced polarization measurements taken at different 
labs (with good angular distributions at only a few energies), $\Sigma$ cross-section 
asymmetry measurements~\cite{Gorbenko:1982ym,Adamian:1991zi,Blanpied:2000xb,Wartenberg:1999}, 
along with a recent set of tensor analyzing powers~\cite{Rachek:2006dz} spanning 25--600 MeV.  
Also, no data of polarization transfer for circularly polarized photons had been taken 
below 500 MeV\@.  We obtained a systematic set of the induced recoil polarization 
observables between 277 and 357 MeV in order to identify where in energy the 
measurements and the existing calculations begin to diverge.  
Benchmark measurements of transferred recoil polarization were also 
taken in this energy region to further constrain the theory.

The experiment was carried out in Hall A of Jefferson Lab~\cite{Leemann:2001dg}.  
A continuous electron beam with longitudinal polarization ranging from 80--85\% 
was produced using a strained gallium-arsenide (GaAs) 
source~\cite{Sinclair:2007ez,HernandezGarcia:2008zz}.  
The longitudinal polarization in Hall A was limited to 38--41\% due to multi-hall running.  
The beam helicity was flipped pseudo-randomly at 30 Hz; beam charge asymmetries between 
the two helicity states were negligible.  The electron beam, with energy 362 MeV, 
was incident on a copper radiator with thicknesses of 3, 4 or 5\% of a radiation length.  
The outgoing (untagged) circularly polarized Bremsstrahlung photons were incident 
on a 15 cm long liquid deuterium target.  The ratio of photon to electron polarization 
varied from 80 to near 100\% and was calculated on an event-by-event basis 
using the formula found in~\cite{Olsen:1959zz}.  

The protons were detected in the left High Resolution Spectrometer (HRS)~\cite{Alcorn:2004sb}, 
made up of one dipole and three quadrupole magnets.   The vertical drift chambers, or VDCs, were 
used to track the protons after the magnetic field of the dipole.  The HRS optics matrix was used 
to reconstruct the scattering angles, momentum and positions at the target.  Triggering 
and time-of-flight information was provided by two planes of plastic scintillators, S1 and S2.  

The Focal Plane Polarimeter (FPP), downstream of the VDCs and trigger panels, is used to
determine the recoil polarization of the protons by measuring a secondary scattering of the protons 
with a $\sim$ 1.7 g/cm$^{3}$ carbon analyzer, 
where the spin-orbit coupling between the proton spin and the orbital angular momentum about the carbon
nucleus leads to an asymmetry in the azimuthal scattering angle, $\phi_{fpp}$.  
GEANT~\cite{Agostinelli:2002hh} Monte Carlo 
studies were used to 
determine the optimal analyzer thickness for each kinematic setting (varied from 0.75'' to 3.75'').
The transferred 
and induced proton polarizations at the focal plane were extracted using the maximum likelihood 
method taking into account the beam helicity state for each event~\cite{Wijesooriya:2001yu}.  
The focal plane polarizations were transported back 
to the target using COSY~\cite{Berz1987402}, a differential algebra based code.  
A detailed description of the polarimeter can be found in~\cite{Punjabi:2005wq,Jones:1997hs}.
 
The experiment covered an angular range of $\theta_{cm}$ = 20$^{\circ}$ -- 120$^{\circ}$, 
generally in 10$^{\circ}$ steps, although some intermediate angles were skipped due to 
time constraints.  Five 20 MeV bins in photon energy spanning 277--357 MeV (bin center) 
were covered at each center of mass angle using two spectrometer momentum settings, except 
at the three largest angles and one of the intermediate ones.  In all measurements the proton 
had sufficient momentum to exclude the existence of a pion in the final state.  Background 
due to electrodisintegration reactions and interactions with the target walls was 
subtracted using a method similar to previous photodisintegration experiments~\cite{Wijesooriya:2001yu}.

The FPP was calibrated with $ep$ elastic scattering~\cite{Ron:2007vr}, which determines 
the false asymmetry and the analyzing power~\cite{Glister:2009cu}, the strength of the 
spin-dependent {\it{p-C}} interaction.  False asymmetries, caused by chamber misalignments 
and inhomogeneities in detector efficiency, cancel to first order for polarization transfer 
but remain for the induced polarization.  The FPP chambers were aligned both internally 
and to the VDCs using straight-through trajectories, with the analyzer block removed.  
The remaining false asymmetries were parameterized as a Fourier series and subtracted out.

\begin{figure*}[ht]
    \centering
    \includegraphics[width=16cm]{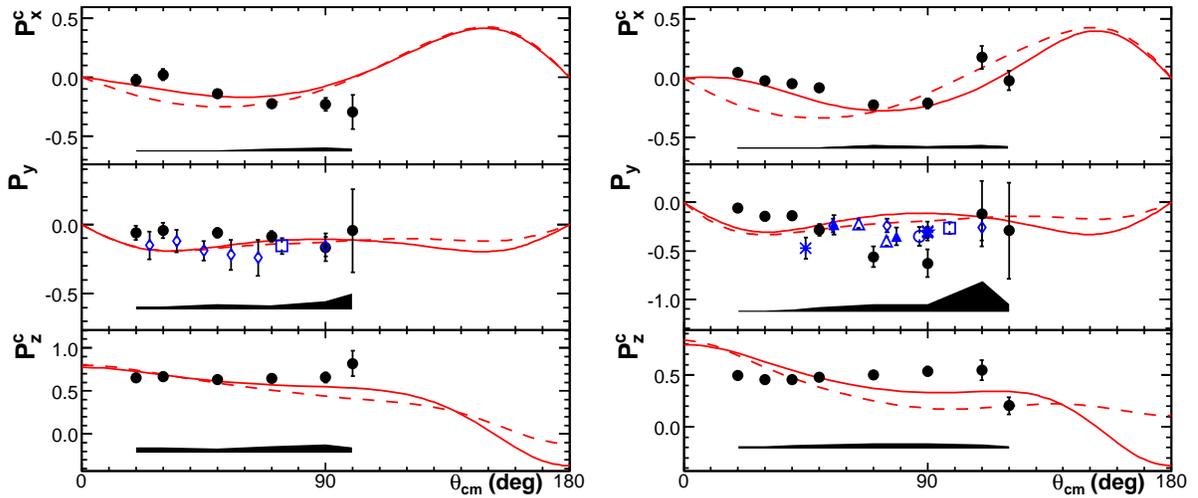}
\caption{(Color online) Angular distributions of the recoil proton polarizations
in $d(\vec{{\gamma}},\vec{p})n$, for $E_{\gamma}$ = 277 $\pm$ 10 MeV (left-hand side) 
and $E_{\gamma}$ = 357 $\pm$ 10 MeV (right-hand side).  Error bars are statistical only; 
systematic uncertainties are shown as black bands.  The solid line is 
the Schwamb and Arenh\"{o}vel calculation~\cite{Schwamb:1999qd,Schwamb:2000zr,Schwamb:2001jy,Arenhoevel:2002kh}.  
The dashed line is a recent improvement from~\cite{Schwamb2010109}.  New data are denoted by the 
filled circles and previous $P_{y}$ data are from Stanford (filled triangles)~\cite{PhysRev.165.1478}, 
Bonn (open squares)~\cite{Kose:1969fn}, Tokyo (open circles)~\cite{Kamae:1978hn}, 
Yerevan (open triangles)~\cite{Avakyan:1990sc} and 
Kharkov (asterisks~\cite{brata80_31,brata80_32,brata86_43}, filled stars~\cite{Barannik:1986ki}, 
open diamonds~\cite{Zybalov1991642}).  Note that there are two overlapping $P_{y}$ measurements 
with different uncertainties for $E_{\gamma}$ = 277 MeV and $\theta_{cm}$ = 90$^{\circ}$.}
\label{fig:prl_figure1}
\end{figure*}

The angular dependence of the new transferred ($P_{x}^{c}$ and $P_{z}^{c}$) and induced ($P_{y}$) 
polarization data are shown as filled circles in Fig.~\ref{fig:prl_figure1} for photon energies 
of 277 MeV (left-hand side) and 357 MeV (right-hand side).  Previous induced polarization measurements 
\cite{PhysRev.165.1478,Kose:1969fn,Kamae:1978hn,brata80_31,brata80_32,brata86_43,Barannik:1986ki,Avakyan:1990sc,Zybalov1991642,Ganenko:1992qa} are also shown, where uncertainty bars are statistical only, except for the 
Tokyo measurements (open circles)~\cite{Kamae:1978hn} which have bars representing 
both statistical and systematic uncertainties.  The uncertainty bars for the 
new measurements are statistical only; systematic uncertainties are shown as black bands.  
The systematic uncertainties include uncertainties in beam energy, polarization and position, 
false asymmetry and analyzing power parameterizations, spin transport, momentum, and 
FPP angular resolution.  The spin transport systematic analysis was similar 
to that of a previous work~\cite{Punjabi:2005wq}.  
 
The solid line is the Schwamb and Arenh\"{o}vel calculation~\cite{Schwamb:1999qd,Schwamb:2000zr,Schwamb:2001jy,Arenhoevel:2002kh}.  
The dashed line is a recent refinement from~\cite{Schwamb2010109}, which includes several technical 
advances such as a non-perturbative treatment of the $\pi NN$ dynamics (as opposed to an approximate treatment).  
The new calculation fulfills unitarity to leading order, requires fewer free parameters and is more rigorous 
from a conceptual point of view, as it considers seven reactions simultaneously in a coupled channel approach, 
rather than only $\gamma d \rightarrow p n$ and $NN\rightarrow NN$.
  
The calculations 
of Schwamb \& Arenh\"{o}vel
described fairly well
the transferred polarizations, $P_{x}^{c}$ and $P_{z}^{c}$.
At the lowest energy, these calculations are also in
good agreement with the 
induced polarization, $P_{y}$, but at the higher energies the 
new data show a strong deviation from the theoretical predictions.
The calculations also appear to overestimate the magnitude of $P_{y}$ for small $\theta_{cm}$, 
even at the lowest energy of 277 MeV\@.  Since $P_{x}^{c}$ and $P_{y}$ are the real and imaginary 
parts of the same sum of amplitudes, the real part appears to be better predicted above 
the $\Delta$ resonance (dominant at roughly 300 MeV) while the imaginary part is better predicted 
at low energy, below the $\Delta$ resonance. 

$P_{z}^{c}$ arises from the combination of amplitudes squared and aside from the lowest energy 
the data exhibit a different angular dependence than both theoretical predictions.  
The disagreement may be due to something being inaccurately approximated 
in the calculations ({\it e.g.}\ the restriction to one-meson exchange).  
Also, free parameters in the potential, which enter into the final-state interactions, 
were also fit to $NN$ scattering data and aside from the dominant $^{1}D_{2}$ channel several 
other partial waves were described only fairly well~\cite{Schwamb:1999qd}. 
Alternatively, QCD-inspired potentials, 
based on chiral perturbation theory (see~\cite{Schwenk:2009wt} for a review), may have 
to be considered.  However, these potentials are currently only considered up 
to the pion threshold, indicating a need for improved QCD-inspired potentials in the $\Delta$-region.
 
\begin{figure}[ht]
\centering
\includegraphics[width=8.5cm]{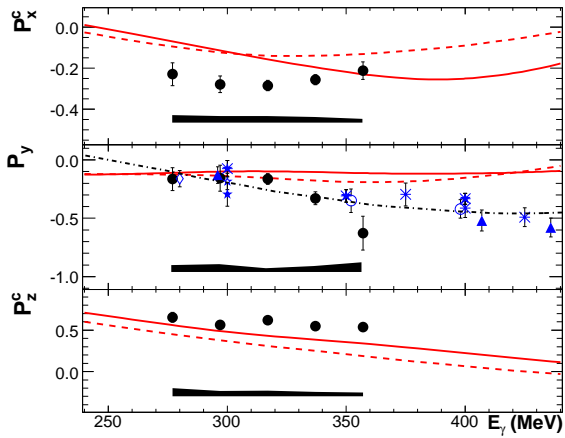}
\caption[width=12cm]{(Color online) Energy distribution of the recoil 
proton polarization in $d(\vec{{\gamma}},\vec{p})n$ for $\theta_{cm}$ = 90$^{\circ}$.  
Error bars are statistical only; systematic uncertainties are shown as black bands.  
The solid line is the Schwamb and Arenh\"{o}vel calculation~\cite{Schwamb:1999qd,Schwamb:2000zr,Schwamb:2001jy,Arenhoevel:2002kh}, 
the dashed line is a recent improvement from~\cite{Schwamb2010109} and the dashed-dotted curve 
is the Kang {\it et al.}\ calculation~\cite{kang1990}.  New data are denoted by the filled 
circles and previous $P_{y}$ data are from Stanford (filled triangles)~\cite{PhysRev.165.1478}, 
Tokyo (open circles)~\cite{Kamae:1978hn} and Kharkov (asterisks~\cite{brata80_31,brata80_32,brata86_43}, 
filled stars~\cite{Barannik:1986ki}, open diamonds~\cite{Zybalov1991642}, open stars~\cite{Ganenko:1992qa}).}
    \label{fig:ledex_E_90deg}
  \end{figure}

The energy dependence at $\theta_{cm} = 90^{\circ}$ is shown in Figure~\ref{fig:ledex_E_90deg} 
for photon energies of 280--480 MeV\@.  The Kang {\it et al.}\ calculation for $P_{y}$ is denoted 
by the dashed-dotted curve and describes in shap of the data in this energy region.
The Schwamb and Arenh\"{o}vel 
calculations (new and old denoted by solid and dashed lines, respectively) predict $P_{y}$ 
to remain relatively flat and near zero below 500 MeV, whereas the new $P_{y}$ results reach 
a value of -0.63 $\pm$ 0.14 $\pm$ 0.09 by 357 MeV\@.  Beyond an energy of roughly 320 MeV, 
the Schwamb and Arenh\"{o}vel calculations are unable to describe the new $P_{y}$ measurements.  
The increasingly poor agreement for $P_{y}$ as the energy is raised may be due to tails of higher 
lying resonances.  Within the impulse approximation, it was found that neither the $D_{13}$ (invariant mass $W$ = 1520 MeV) 
or $S_{11}$ ($W$ = 1535 MeV) resonances played significant roles in $P_{y}$ below 400 MeV~\cite{Schwamb:1995fh}.  
However, a coupled channel approach involving the $D_{13}$ and $S_{11}$ as well as the Roper 
resonance ($P_{11}$ with $W$ = 1440 MeV) and possibly a double $\Delta$ 
excitation (both of which were included approximately within the impulse approximation in~\cite{Schwamb:1995fh}) 
may have to be considered.

Note that the older and new calculations are roughly equivalent, aside 
from a divergence at the highest energies for the polarization transfers, where the 
older calculations more closely resemble the data.  The new calculation (dashed line) 
gives a slightly better description of $P_{y}$ while the older calculation (solid line) 
more accurately describes $P_{z}^{c}$.  A similar situation was observed~\cite{Rachek:2006dz} 
for the deuteron photodisintegration tensor analyzing powers ($T_{20}$, $T_{21}$ and $T_{22}$), 
where it was found that the new calculation was better at describing $T_{20}$ and $T_{22}$ 
but worse at describing $T_{21}$.  Neither the old or new calculations describe the energy 
dependence of $P_{x}^{c}$ at $\theta_{cm}$ = 90$^{\circ}$, except for the agreement at 
357 MeV between the new data and the older calculation.  The overall shape of $P_{z}^{c}$ 
as a function of energy appears to be modeled well, but the magnitude is underestimated by 
both calculations (more-so by the new model).

To summarize, we have provided new induced and transferred recoil proton polarization measurements 
for deuteron photodisintegration over a wide range of energies and angles.  The new induced 
polarization measurements are consistent with theoretical predictions and previous measurements 
at all but the highest energies, where the most technically advanced MBM calculations appear 
unable to describe the large value of $P_{y}$.  $P_{x}^{c}$ appears to be described well using 
an older MBM calculation at the highest energy while $P_{z}^{c}$ agrees with the same 
calculation (within uncertainties) at the lowest energy.  It may be possible to remedy 
the situation by improving the fits of $NN$ scattering partial waves, including higher 
mass resonances in a coupled channel approach or extending the calculation beyond the 
one-meson approximation.  These new measurements should provide input for important tests to the state-of-the-art meson-baryon model calculations above pion threshold and it will be interesting to see whether the issue can be resolved or if other 
models (based on chiral perturbation theory, for instance) should be considered.

We thank the Jefferson Lab physics and accelerator divisions for their contributions.  
This work was supported by the U.S. Department of Energy (including contract DE-AC02-06CH11357), 
the U.S. National Science Foundation, the Israel Science Foundation, the Korea Science Foundation, 
the US-Israeli Bi-National Scientific Foundation, the Natural Sciences and Engineering Research 
Council of Canada, the Killam Trusts Fund, the Walter C. Sumner Foundation and the Deutsche 
Forschungsgemeinschaft (SFB 443).  Jefferson Science Associates operates the Thomas Jefferson 
National Accelerator Facility under DOE contract DE-AC05-06OR23177.  The polarimeter was 
funded by the U.S. National Science Foundation, grants PHY 9213864 and PHY 9213869.

\begin{singlespace}
\bibliography{ledex}
\end{singlespace}

\end{document}